\documentclass[journal]{IEEEtran} 
\usepackage{array}
\usepackage[tight,footnotesize]{subfigure}
\usepackage{bm}
\usepackage{amsfonts}
\usepackage{amsmath}
\usepackage{epsfig}
\def\Tr{{\rm Tr}}
\def\MSE{{\rm MSE}}
\def\SNR{{\rm SNR}}
\def\MMSE{{\rm MMSE}}

\def\la{\left\langle}
\def\ra{\right\rangle}

\def\bX{{\bar X}}
\def\bfX{{\bf X}}
\def\bfx{{\bf x}}
\def\bfY{{\bf Y}}
\def\bfy{{\bf y}}
\def\bfU{{\bf U}}
\def\bfu{{\bf u}}
\def\bE{\mathbb{E}}
\def\bnabla{\bm{\nabla}}
\def\blambda{\bm{\lambda}}
\def\bJ{{\bf J}}
\def\bC{{\bf C}}

\def\be{\begin{equation}}
\def\ee{\end{equation}}
\def\ber{\begin{eqnarray}}
\def\eer{\end{eqnarray}}
\begin{document}
\title{\bf Certain Relations between Mutual Information and Fidelity of Statistical Estimation}
\author{Sudhakar~Prasad,~\IEEEmembership{Fellow,~OSA and Member,~SPIE}%
\thanks{The author is with the Department of Physics
and Astronomy, University of New Mexico, Albuquerque, New Mexico 87131}
\thanks{Manuscript submitted to IEEE Trans. Inform. Th., October 6, 2010}}

\maketitle
\begin{center}
This work has been submitted to the IEEE for possible
publication. Copyright may be transferred
without notice, after which this version may no longer be accessible.
\end{center}
\begin{abstract}

I present several new relations between mutual information (MI) and statistical estimation error for
a system that can be regarded simultaneously as a communication channel and as an estimator of an input parameter.
I first derive a second-order result between MI and Fisher information (FI) that is valid
for sufficiently narrow priors, but arbitrary channels. 
A second relation furnishes a lower bound on the MI in terms of 
the minimum mean-squared error (MMSE) 
on the Bayesian estimation of the input parameter from the channel output, one
that is valid for arbitrary channels and priors.
The existence of such a lower bound, while extending previous work relating the MI to
the FI that is valid only in the asymptotic
and high-SNR limits, elucidates further the fundamental connection between
information and estimation theoretic measures of fidelity.
The remaining relations I present are inequalities and correspondences among MI, FI, and   
MMSE in the presence of nuisance parameters.
\end{abstract}
\IEEEpeerreviewmaketitle

\section{Introduction}\label{sec:intro}

Statistical information theory \cite{Shannon48,CoverThomas91} constitutes an essential tool for 
modern signal processing, computation, coding, and communication systems.
Its core philosophy hinges on the notions of information potential and
the ability of systems to encode, transmit and decode information
about one or more input parameters.

Based in statistical estimation theory, Fisher information (FI) \cite{vanTrees68}
on the other hand represents the sensitivity of statistical data to 
one or more input parameters. Its inverse, the so-called Cram\'er-Rao bound (CRB),
yields a useful lower bound on the variance of any statistical data-based estimation
of those parameters.

In spite of the different essential motivations for the two families of information measures, 
mutual information (MI) and FI 
are closely related at least asympotically in the limit of a large number of conditionally
independent measurements \cite{ClarkeBarron90,Rissanen96,BrunelNadal98,KangSompolinsky01}.
A recent paper explores the validity of this asymptotic relationship when the 
number of measurements is not particularly large \cite{ChallisYarrowSeries08}.

The relation of MI to FI is essentially a local one that is valid only in the limit of either a narrow channel PDF, 
as in Ref.\cite{ClarkeBarron90}, or a narrow input PDF, as we shall see in this paper.  
In the more general case, 
the MI, as I shall also show, is related more naturally to the minimum mean squared error (MMSE) 
of Bayesian estimation. Unlike previous work \cite{Duncan70,GuoShamaiVerdu05,WolfZakai07},
on this topic, the new relation, a lower bound on the MI, is general and applicable
to arbitrary channel and input statistics. It may be regarded as a
global generalization of the more restrictive local relations between MI and FI.

A number of additional correspondences between the MI and MMSE are derived that apply
when either more measurements, or channels, are added or multiple input parameters must be estimated
at once. In the latter case if the input parameters are statistically independent, then each parameter serves 
as a nuisance for the other parameters that must, in general, reduce both the MI and the fidelity
of estimation for each parameter.
These local and global considerations on the fundamental relationship between MI and Bayesian estimation error
are the subject of this paper.

\section{A Second-Order Relation between MI and FI} \label{sec:problem}

Let $X$ be an input parameter that is
statistically distributed according to the probability density function (PDF)
$P(x)$ \cite{footnote1} with mean $\bar X$ and variance $\sigma_X^2$. 
Let $Y$ be an output variable, {\it e.g.,} a measurement variable, that carries information 
about $X$, and is distributed according to the PDF $P(y)$.
For notational definiteness, let us take these variables to be continuous over appropriate ranges of values, 
but the analysis of this section applies 
equally well to discrete random variables too, provided all integrals over such variables are
regarded as discrete sums over the corresponding sample spaces.

The communication channel, or the measurement system as the case may be, is described
by means of the conditional PDF, $P(y|x)$. In spite of the notation, there is no restriction
placed on the number of output variables represented by the symbol. In other words, $Y$ is
in general a multi-dimensional output vector. Although I shall for clarity assume initially that
the input is one-dimensional, the generalization to multiple input parameters,
as we shall see subsequently, is straightforward. 

The three PDFs are related according to the Bayes' rule,
\begin{equation}
\label{e1}
P(y) =\int P(y|x)\, P(x)\, dx.
\end{equation}
The MI is defined in terms of the various PDFs by three different entirely equivalent expressions,
\begin{align}
\label{e2}
I(X;Y)&= h(X)-h(X|Y)\nonumber\\
      &= h(Y)-h(Y|X)\nonumber\\
      &= h(X)+h(Y)-h(X,Y),
\end{align}
where for each PDF $h$ denotes the corresponding differential entropy defined by averaging 
the negative logarithm of the PDF over the joint PDF, $P(x,y)$,
\begin{align}
\label{e3}
h(X)&= -\int P(x)\, \ln P(x)\, dx;\nonumber\\
h(X|Y)&= -\iint P(x,y)\, \ln P(x|y)\, dx\, dy;\nonumber\\
h(X,Y)&= -\iint P(x,y)\, \ln P(x,y)\, dx\, dy;
\end{align}
and so on.
I shall always use the natural logarithm for the definition of entropies
in this paper, as it yields the simplest 
form of the final results. All entropy and information measures are thus 
expressed in natural units, or nats.

By using definitions of form (\ref{e3}) in the second of the expressions (\ref{e2}) and using
the Bayes relation (\ref{e1}), we may express the MI as the average
\begin{equation}
\label{e4}
I(X;Y) = -\bE \left[\ln\, \int P(x^\prime) {P(y|x^\prime)\over P(y|x)} dx^\prime\right].
\end{equation}
By expanding $P(y|x^\prime)$ in a Taylor series of powers of the
deviation $(x^\prime-x)$, we may transform the logarithmic term in Eq.~(\ref{e4}),
\begin{equation}
\label{e5}
\ln \int P(x^\prime) {P(y|x^\prime)\over P(y|x)} dx^\prime =\ln \left[1+\sum_{n=1}^\infty {\sigma^{(n)}(x)
\over n!P(y|x)} {\partial^n P(y|x)\over \partial x^n}\right],
\end{equation}
where the $x$ dependent ``moments" of the $X$-PDF are defined as
\begin{equation}
\label{e6}
\sigma^{(n)}(x)=\int P(x^\prime) \, (x^\prime -x)^n \, dx^\prime.
\end{equation}
By subtracting $\bar X$, the mean value of $X$, from both $x^\prime$ and $x$ inside the integrand
in Eq.~(\ref{e6}) and noting that linear deviations from the mean average to 0, 
we may easily evaluate the first two $x$-dependent moments as
\begin{equation}
\label{e7}
\sigma^{(1)}(x)=-(x-\bar X);\ \ \sigma^{(2)}(x) = \sigma_X^2 +(x-\bar X)^2. 
\end{equation}

We may now expand the logarithm (\ref{e5}) to second order in the deviations
and note that 
\begin{equation}
\label{e8}
\int dy\, P(y|x){1\over P(y|x)} {\partial^n P(y|x)\over \partial x^n}= {d^n\over dx^n}\int P(y|x) \, dy = 0
\end{equation}
for all $n\geq 1$. In view of this result, the only contributing term to the second order
is $-(1/2)\left[\sigma^{(1)}(x)\right]^2 (\partial \ln \, P(y|x)/\partial x)^2$. Substituting this 
term into Eq.~(\ref{e4}) yields to the second order the following expression for the MI, $I(X;Y)$:
\begin{align}
\label{e9}
I(X;Y) &= {1\over 2}\int dx\,  P(x)\, \left[\sigma^{(1)}(x)\right]^2 \, \int dy\, P(y|x) 
\nonumber\\
&\qquad\times \left[{\partial \ln \, P(y|x) \over \partial x}\right]^2\nonumber\\
&= {1\over 2}\int dx\,  P(x)\, (x-\bar X)^2 \, J(Y|x),
\end{align}
where $J(Y|x)$ is the FI defined locally at each value of $X$ as
\begin{equation}
\label{e10}
J(Y|x) = \int dy\, P(y|x) \left[{\partial \ln \, P(y|x) \over \partial x}\right]^2.
\end{equation}

This is the first important result of the paper. Its validity is guaranteed for sufficiently narrow
priors for which the higher-order deviations about the input mean are negligible.
Note the non-local character of this second-order equality (\ref{e9}): The MI is a squared-deviation-weighted average
of the FI, the latter evaluated locally over the full sample space of $X$. 

For multiple-input, multiple-output (MIMO) channels, the following multi-parameter analog 
of the second-order result (\ref{e9}) is easily derived as well:
\begin{equation}
\label{e11}
I(X;Y) = {1\over 2}\int d{\bf x}\,  P({\bf x})\, \sum_{j,k}\delta x_j\delta x_k J_{jk}(Y|{\bf x}),
\end{equation}
where $\delta x_j\equiv x_j-\bar X_j$ denotes the deviation of the $j$th component of the input vector ${\bf x}$
from its mean value.

It is also possible to extend relations (\ref{e9}) and (\ref{e11}) to the case of {\em discrete} random input parameters
by replacing all integrals over $x$ to discrete sums over values in the sample space of $X$, writing instead of Eq.~(\ref{e5})
\begin{equation}
\label{e11a}
-\ln \bE_{X} {P(y|X)\over P(y|x)}=-\ln \left\{1+\bE_X\left[{P(y|X)-P(y|x)\over P(y|x)}\right]\right\},
\end{equation}
expanding the logarithm in a power series, and then noting that up to the second order it may be expressed as
\begin{align}
\label{e11b}
{1\over 2}&\left\{\bE_X\left[{P(y|X)-P(y|x)\over P(y|x)}\right]\right\}^2\nonumber\\
&\leq {1\over 2}\bE_X\left\{\left[
{P(y|X)-P(y|x)\over P(y|x)}\right]^2\right\},
\end{align}
where the inequality follows from a simple application of the Cauchy-Schwarz inequality.
The subscript $X$ to the expectation-value symbol indicates that the expectation is
taken relative to $X$, keeping other variables fixed.
An expectation of the RHS above, first over $y$, given $x$, and finally over $x$ yields
the following upper bound on MI to the second order:
\begin{equation}
\label{e11c}
I(X;Y)\leq {1\over 2}\bE_{X}\bE_{X^\prime} [K(X,X^\prime)],
\end{equation}
where $K(X,X^\prime)$ defined by
\begin{equation}
\label{e11d}
K(X,X^\prime)\equiv \bE \left\{\left[{P(Y|X^\prime)-P(Y|X)\over P(Y|X)}\right]^2\Bigg| X,X^\prime\right\}
\end{equation}
is the Chapman-Robbins information (CRI) \cite{ChR51}. For a fixed value of $X$, the CRI when minimized over all possible 
values that $X^\prime$ can take yields, via its reciprocal, the tightest lower bound on
the error in estimating the discrete variable in the single-test-point optimization subspace.
Note that the upper bound (\ref{e11c}) applies to MIMO channels as well.

The results of this section have a simple interpretation: For a narrow input PDF, the MI, like the FI and
CRI, is a local sensitivity based
measure of information. The more sensitive the channel PDF -- and thus the data -- to the input, 
the larger all these information measures. 
The Gaussian linear channel illustrates this point well.

\subsection{The Gaussian Linear Channel}
Consider the Gaussian linear channel in which $X$ and $Y$ are related through a linear gain parameter $a$,
a linear bias $b$, and an additive noise $N$ distributed according to a zero-mean Gaussian 
PDF of variance $\sigma_N^2$:  
\be
\label{e12}
Y=aX+b+N, \ \ N\sim {\cal N}(0,\sigma_N^2).
\ee
In this case, the FI of $Y$, given $X=x$, is easily computed to be
\be
\label{e12p}
J(Y|x)={a^2\over \sigma_N^2},
\ee
independent of $x$. In view of this result, the second-order equality (\ref{e9}) becomes $1/2$ times the power SNR,
which is the ratio of $a^2$ times the $X$-variance and the noise variance, 
\be
\label{e13}
I(X;Y)= {1\over 2} {a^2\sigma_X^2\over \sigma_N^2}= {1\over 2} {\rm SNR}.
\ee
Note that the Gaussian-channel result (\ref{e13}) is independent of the statistics of $X$. 
It is also in agreement with the well known expression for the MI of a Gaussian channel with a Gaussian input PDF, 
\begin{equation}
\label{e13a}
I(X;Y)={1\over 2} \ln\, \left(1+{a^2\sigma_X^2\over \sigma_N^2}\right),
\end{equation}
when the latter is expanded to the lowest order in $\sigma_X^2$.

For input PDFs that have arbitrary width, a different relation between the estimation error and
MI can be obtained. The precise relation in this case involves the minimum mean-squared error 
of Bayesian estimation and provides a lower bound on the MI. I next derive this lower bound.

\section{Bayesian Estimation and Minimum Mean-Squared Error} 

A good Bayesian estimation is one that reduces the mean squared error (MSE) to a value below the 
variance of the input PDF, the so-called prior.
The variance of the prior represents the maximum MSE incurred by electing to use the mean of the prior as
the trivial estimator when no information from data is availaible as, e.g., in the limit of a vanishing SNR.

The MSE of a Bayesian estimator, $\hat X(Y)$, of $X$ is defined as
\be
\label{e14}
\MSE_{\hat X}=\bE\left\{ [\hat X(Y)-X]^2\right\},
\ee
where the statistical average is taken over the joint distribution of $X$ and $Y$. 
The estimator $\hat X$ that minimizes the MSE
is called the minimum-MSE estimator (MMSEE) \cite{Kay93}. It is easily shown to be the 
mean of $X$, given $Y$, {\it i.e.,} its {\it posterior} mean,
\be
\label{e15}
{\hat X}_M(Y)=\bE (X|Y) =\int x \, P(x|Y)\, dx.
\ee
Its mean value is the mean of the prior, $\bar X$.

The MSE corresponding to the MMSEE is the minimum MSE (MMSE) that provides the
tightest possible lower bound for the MSE of {\it any} Bayesian estimator of $X$. Since
$[\hat X(Y)-X]^2 = \hat X^2(Y)-2X\hat X+X^2$, we may express the MSE (\ref{e14}) for the MMSEE, {\em i.e.,}
the MMSE as
\begin{align}
\label{e16}
\MMSE&=\bE(X^2) -2\bE[\bE(X|Y)\hat X_{M}(Y)]+\bE [\hat X_{M}^2(Y)]\nonumber\\
&=\bE(X^2)-\bE[\hat X^2_{M}] = \sigma_X^2-\sigma_{M}^2,
\end{align}
where the last two equalities are obtained by recognizing that $\bE(X|Y)$ is the MMSEE, $\hat X_M(Y)$, 
and that $X$ and $\hat X_{M}$ both have the same expectation.
Since variance is always non-negative, the last equality proves that the MMSE
can never exceed the prior variance.

\section{Relation between Mutual Information and MMSE}

The conditional differential entropy, $h(X|Y)$, sometimes called equivocation, may be expressed
as a statistical average over the output, $Y$,   
\be
\label{e17}
h(X|Y) = -\bE\left[\int P(x|Y)\, \ln P(x|Y)\, dx\right]
\ee
where the argument of the $Y$-average is the conditional entropy, given a fixed value of $Y$.
But for a given variance, $\sigma_{X|Y}^2$, of the PDF $P(x|Y)$, its entropy is
bounded above by the entropy of a Gaussian PDF with the same variance \cite{CoverThomas91},
namely $(1/2)\ln(2\pi e\sigma_{X|Y}^2)$.
As a result, the conditional differential entropy (\ref{e18}) is
bounded above as follows:
\begin{align}
\label{e18}
h(X|Y) &\leq {1\over 2}\bE_Y\left[ \ln\left(2\pi e\sigma_{X|Y}^2\right)\right]\nonumber\\
&\leq
{1\over 2}\ln(2\pi e)+{1\over 2}\ln \left[\int dy \, P(y)\,\sigma_{X|Y}^2\right],
\end{align}
where the second inequality results from the convexity of the logarithm.

To see that the integral on the RHS of the second of the relations (\ref{e18}) evaluates 
to the MMSE, we may note that in view of relation (\ref{e15})
\begin{align}
\label{e19}
\sigma_{X|Y}^2 &=\int \left\{x-\left[\int x\, P(x|Y)\, dx\right]\right\}^2 P(x|Y)\, dx\nonumber\\
 &=\int \left[{\hat X}(Y)-x\right]^2 P(x|Y) \, dx,
\end{align}
whose $Y$-average is simply the MSE for the MMSEE estimator, namely the MMSE.
(To simplify notation here and in the rest of the paper, 
I have omitted the subscript $M$ from the MMSE estimator.)
Putting results (\ref{e18}) and (\ref{e19}) together, we arrive at the following upper bound on equivocation:
\be
\label{e20}
h(X|Y) \leq {1\over 2} \ln\left( 2\pi e\, \MMSE\right)
\ee
and the corresponding lower bound on the MI (\ref{e2}):
\be
\label{e21}
I(X;Y) \geq h(X)-{1\over 2} \ln\left( 2\pi e\, \MMSE\right).
\ee

Result (\ref{e21}) is the second major contribution of this paper.  
It demonstrates the precise inverse relationship between the {\it minimum} Bayesian estimation error and 
the {\it minimum} statistical information that can be transmitted by the measurement channel. 
For an additive, linear Gaussian channel with a Gaussian input, both inequalities in Eq.~(\ref{e18}) become equalities,
the first because in this case $P(X|Y)$ is Gaussian and the second because $\sigma_{X|Y}^2$
is independent of $Y$. Consequently, for such channel and input, the inequality (\ref{e21}) is
obeyed as an equality.  
Indeed, since the MMSE for this case is simply $\Big(\sigma_X^{-2}+a^2\sigma_{Y|X}^{-2}\Big)^{-1}$,
while $h(X)$ is $(1/2)\ln (2\pi e\sigma_X^2)$, we have the well known result, $(1/2)\ln (1+\SNR)$,
for MI, where $\SNR=a^2\sigma_X^2/\sigma_{Y|X}^2$ is the power SNR and 
$a$ is the linear gain factor of the Gaussian channel.
The derivative equality obtained in \cite{GuoShamaiVerdu05},
\be
\label{e22}
{d\over d \SNR} I(X;Y) ={1\over 2\sigma^2_X}\MMSE,
\ee
is a simple, immediate consequence of this result specific to Gaussian channels.

For a non-Gaussian channel, the lower bound (\ref{e21}) on the MI, $I(X;Y)$, is in general
not attainable. I now analyze the Poisson channel with a negative-exponential prior 
to illustrate this fact. 

\subsection{The Linear Poisson Channel with a Negative-Exponential Prior}

Consider the linear Poisson channel with linear gain (or, scaling) factor $a$ and linear bias $b$, so
the conditional mean of output $Y$, given input $X$, is $\bE( Y|X) = a X + b$. 
The conditional Poisson probability distribution (PD) over the discrete samples of $Y$, given $X$, has the form
\be
\label{e23}
p(y|x) ={(ax+b)^y\over y!} \exp[-(ax+b)],\ \ y=0,1,2,\ldots .
\ee
If we take the prior PDF to be negative exponential with mean $\bar X$,
\be
\label{e24}
P(x) =\left\{\begin{array}{ll}
{1\over \bar X}\exp(-x/\bar X) & {\rm for}\  x\geq 0\\
0 & {\rm otherwise,}
\end{array}\right.
\ee
then by Bayes theorem the unconditional $Y$-PDF takes the form
\begin{align}
\label{e25}
p(y) =\int_0^\infty dx {(ax+b)^y\over y!}& \exp[-(ax+b)]\exp(-x/\bar X),\nonumber\\
& y=0,1,2,\ldots .
\end{align}
By a suitable scaling and shift of the integration variable, this integral may be expressed in 
terms of the incomplete Gamma function,
\be
\label{e26}
\Gamma(y+1,u)=\int_u^\infty dx\, \exp(-x)\, x^y,
\ee
as
\be
\label{e26}
p(y)={1\over y!}{(a\bar X)^y\over (a\bar X+1)^{y+1}}\exp(b/a\bar X)\Gamma(y+1,b(a\bar X+1)/a\bar X).
\ee

The following expression for the mean squared MMSEE, $\bE [\hat X^2(Y)]$, 
is a simple consequence of the definition (\ref{e15}) and the Bayes theorem: 
\be
\label{e27}
\bE\left[ \hat X^2(Y)\right] = \sum_{y=0}^\infty {K(y)^2\over p(y)},
\ee
where $K(y)$ denotes the expression
\be
\label{e28}
K(y)=\int dx\, x\, P(x) \, p(y|x).
\ee
For the Poisson channel and negative-exponential prior, $K(y)$ may be expressed in terms of $p(y)$,
since the latter has a similar expression as (\ref{e28}) with the only difference that 
the factor $x$ is missing from the integrand. To see this, we first write $x=(1/a)(ax+b)-b/a$ in 
expression (\ref{e28}) and then recognize that for the Poisson channel PD given by Eq.~(\ref{e23})
$(ax+b) p(y|x)$ equals $(y+1)$ times $p(y+1|x)$. This yields the following useful form for $K(y)$:  
\be
\label{e29}
K(y)= {(y+1)\over a} p(y+1) - {b\over a} p(y).
\ee
Substituting this expression into Eq.~(\ref{e27}) and noting that 
\be
\label{e30}
\sum_{y=0}^\infty (y+1)\, p(y+1) = \la Y\ra = a\bar X +b;\ \sum_{y=0}^\infty p(y)= 1;
\ee
and $E(X^2) = 2{\bar X}^2$ for the NE prior (\ref{e24}), we obtain the following 
expression for the MMSE (\ref{e16}):
\be
\label{e31}
\MMSE= 2{\bX}^2+2{b\over a} \bX +{b^2\over a^2}-{1\over a^2}\sum_{y=0}^\infty {(y+1)^2\,p^2(y+1)\over p(y)}.
\ee
We can now numerically evaluate the MMSE expression (\ref{e31})in the general case of arbitrary $a$ and $b$,
but for the case of zero bias, $b=0$, a simple analytical expression can be derived as we now show. 

\subsubsection{The Case of Zero Bias, $b=0$}

Expression (\ref{e26}) for $p(y)$ now greatly simplifies since the incomplete Gamma function
in that expression becomes complete, taking the value $y!$, and the sum in expression (\ref{e31})
may now be easily performed analytically, since
\begin{align}
\label{e32}
\sum_{y=0}^\infty {(y+1)^2\,p^2(y+1)\over p(y)} &= {a\bX\over (a\bX + 1)^2}\sum_{y=0}^\infty (y+1)^2\alpha^{y+1},
\nonumber\\
&\qquad \alpha\equiv {a\bX \over a\bX+1},
\end{align}
is related to the sum $\sum_{y=0}^\infty \alpha^y=(1-\alpha)^{-1}$ by two successive applications of the differential operator,
$\alpha\, {\partial/\partial\alpha}$. This yields the following simple expression for the MMSE when $b=0$:
\be
\label{e33}
\MMSE= {\bX^2\over 1+a\bX}.
\ee
This expression has the desired property of reducing to the prior variance, $\bX^2$, in the limit of vanishing SNR, $a\bX \to 0$, and
of vanishing in the opposite limit, $a\bX \to\infty$.

The MI may also
be evaluated for the Poisson channel and negative exponential prior, most simply via the second of the expressions (\ref{e2}).
Since $\ln\, p(y|x)= y\, \ln(ax+b)-(ax+b)-\ln y!$, the conditional mean of $-\ln p(y|x)$, given $x$, is simply 
\be
\label{e34}
-\bE [p(Y|x)] = -(ax+b)[\ln(ax+b)-1] +\bE_{Y|x} (\ln Y!).
\ee
A subsequent average over the prior $P(x)$ then yields the conditional (discrete) entropy $H(Y|X)$, which when
subtracted from the unconditional output entropy $H(Y) =-\bE [\ln p(Y)]$ produces the following exact 
expression for the MI:
\begin{align}
\label{e35}
I(X;Y)&= \int_0^\infty dx \, P(x)\, (ax+b)[\ln(ax+b)-1] \nonumber\\
&-\sum_{y=0}^\infty p(y)\,  \ln [p(y)\, y!].
\end{align}
This too can be evaluated numerically. 

The differential entropy of the negative exponential prior takes a simple analytical form, since $-\ln P(x)=
\ln\bX + x/\bX$ whose mean, the differential entropy of $X$, is simply $1+\ln\bX$:
\be
\label{e36}
h(X)= 1+\ln \bX.
\ee
Use of this expression and the MMSE (\ref{e31}) yields the lower bound (\ref{e21}) on the MI. I now compare
this lower bound numerically with the exact value given by the expression (\ref{e35}).

\begin{figure}
\centerline{
\includegraphics[width=3in]{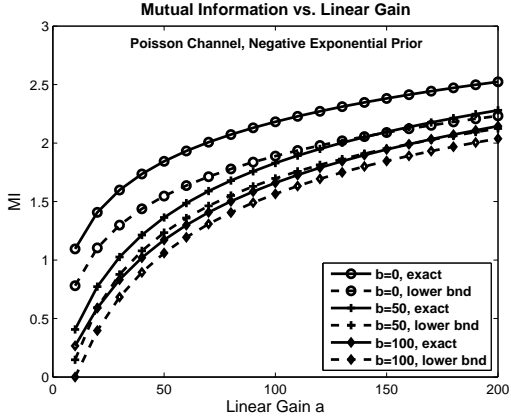}}
\caption{\label{fig:f1} MI vs. the normalized linear gain parameter, $a\bX$, for three different values of
$b$, as indicated. The solid curves refer to the exact result (\ref{e35}),
while the corresponding dashed curves refer to the lower bound (\ref{e21}). }
\end{figure} 

In Fig.~1 I display, as a function of the normalized linear gain parameter $a\bX$, the exact expression (\ref{e35}) 
(solid curves) along with the corresponding lower bound (\ref{e21})
(dashed curves) for three different values of $b$, namely 0, 50, and 100.
The lower bound becomes tighter as the gain parameter increases in value, but typically it fails to provide
a useful, nontrivial lower bound below a certain threshold value of the gain. Indeed, as the
exact expression for the lower bound in the case $b=0$ obtained from Eqs.~(\ref{e36}), (\ref{e33}),
and (\ref{e21}), namely
\begin{align}
\label{e37}
I(X;Y)&\geq h(X)-{1\over 2}\ln(2\pi e\, \MMSE) \nonumber\\
&={1\over 2}\left[1-\ln \left(2\pi\cdot {\MMSE\over \bX^2}\right)\right]
\end{align}
shows, the lower bound drops below the trivial lower bound of 0 for $a\bX$ below $2\pi/e-1\approx 1.31$. A similar
but higher threshold below which the lower bound (\ref{e21}) ceases to be nontrivial is obtained when
$b$ is non-zero. However, as $b$ increases this lower bound becomes increasingly tighter and thus more useful
at sufficiently large values of the normalized gain, $a\bX$.  

\begin{figure}
\centerline{
\includegraphics[width=3in]{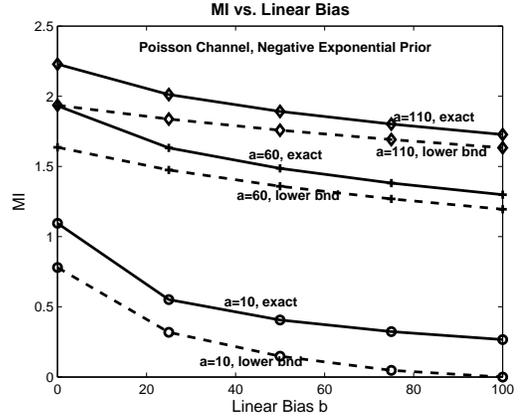}}
\caption{\label{fig:f2} MI vs. the linear bias parameter, $b$, for three different values of
gain $a\bX$, as indicated. The solid curves refer to the exact result (\ref{e35}),
while the corresponding dashed curves refer to the lower bound (\ref{e21}). }
\end{figure} 

In Fig.~2, I plot the exact values and the corresponding lower-bound values for the MI as a function of
the linear bias parameter, $b$, for three different values of the gain parameter, $a\bX$. As $b$ increases,
the MI decreases as expected since the sensitivity of data on the input variable $X$ is reduced.  
Raising the linear gain raises the MI, as expected, for each $b$ value, as the previous figure shows. 
Again, it is clear that the lower bound (\ref{e21}) is useful one for sufficiently large values of $a\bX$ and $b$.

\section{Generalization to Multiple-Input, Multiple-Output Channels}

When MIMO channels are involved, we may organize the input and output variables
into two different column vectors, say $\bfX =(X_1,\ldots,X_N)^T$ and $\bfY = (Y_1,\ldots,Y_M)^T$, where $T$
denotes a matrix transpose.
The $N$-parameter analog of the upper bound (\ref{e18}) is simply 
\begin{equation}
\label{m23}
h(X|Y) \leq {1\over 2}\bE_Y\left\{ \ln\left[(2\pi e)^N\, |\bC_{X|Y}|\right]\right\},
\end{equation}
where $|\bC_{X|Y}|$ denotes the determinant of the positive semi-definite covariance matrix of 
$\bfX$, given $\bfY$. The determinant of such a matrix is a product of its $N$ non-negative eigenvalues,
or simply the $N$th power of their geometric mean. Since the latter cannot exceed the arithmetic
mean of these eigenvalues, which is $1/N$ times the trace of the matrix, and since the logarithm is a convex
function, we have the following inequalities for $h(X|Y)$: 
\begin{align}
\label{m24}
h_{X|Y} &\leq {N\over 2} \ln(2\pi e)\, +\,{N\over 2}\bE\left[ \ln\left({1\over N}\Tr\, \bC_{X|Y}\right)\right] \nonumber\\
 &\leq {N\over 2} \ln(2\pi e/N)\, +\,{N\over 2} \ln\, \bE\left(\Tr\, \bC_{X|Y}\right) \nonumber\\
 &= {N\over 2} \ln(2\pi e \MMSE),
\end{align}
where $\MMSE$ here is the average minimum MSE of a component-wise estimation of $\bfX$, 
\begin{align}
\label{m25}
\MMSE &= {1\over N}\bE_Y \, \bE\left \{[\bfX-\hat\bfX(\bfY)]^T[\bfX-\hat \bfX(\bfY)]\Big|Y\right\}\nonumber\\
&={1\over N}\bE_Y [\Tr\,\bC_{X|Y}], 
\end{align}
involving the MMSEE $\hat\bfX(\bfY)$ for the MIMO problem,
\begin{equation}
\label{m26}
\hat\bfX(\bfY) = \int \bfX \, P(\bfX|\bfY)\, d\bfX.
\end{equation}
Correspondingly, the MI is lower bounded by 
\be
\label{m27}
I(\bfX;\bfY) \geq h(\bfX)-{N\over 2} \ln\left( 2\pi e\, \MMSE\right).
\ee
Note that in the MIMO case each component of the MMSEE minimizes the MSE of the corresponding input parameter, the one it estimates. 
As such, the MMSEE vector (\ref{m26}) as a whole also minimizes the average MSE per component of the input vector.

\section{Additional Properties of MMSE and Their Correspondence with Information}

I now establish two additional important properties of the MMSE not previously reported in the literature but
which help strengthen the correspondences with information I have already discussed via relations (\ref{e10}), (\ref{e11}), (\ref{e21}),
and (\ref{m27}).
The first of these concerns the behavior of the MMSE as additional measurements are made. It is well known
\cite{CoverThomas91,vanTrees68} that both MI and FI exhibit an additive property, namely
\begin{align}
\label{e38}
I(X;Y,Z) &= I(X;Y) + I(X;Z|Y)\geq I(X;Y)\nonumber\\
J(Y,Z;X) &= J(Y;X)+J(Z|Y;X) \geq J(Y;X),
\end{align}
which represents the fact that in general an additional measurement only increases information. The conditional information,
either $I(X;Z|Y)$ or $J(Z|Y;X)$, is a direct measure of the capacity of the measurement $Z$ to improve information
about $X$, given that the measurement $Y$ has already been made. 
Since the estimation variance is lower bounded by the inverse of the FI\footnote{ 
When multiple inputs are involved, the additivity and inequality relations for the FI
as well as its inverse must be interpreted in the matrix sense.}, the two relations (\ref{e38})
represent a useful inverse relationship between MI and estimation error.

But this fundamental relationship is at best a local one since, as I have argued before, the FI and its inverse, the Cram\'er-Rao lower bound
on estimator variance, are local measures of information and estimation fidelity.
I now show that MMSE exhibits a similar behavior, which will serve to accord a general global character to this local inverse relationship 
between information and error. 

\subsection{MMSE Cannot Increase with Measurement}

\def\defeq{\stackrel{\rm def}{=}}
Let us consider two measurements, $Y$ and $Z$, of the input parameter $X$. 
The joint MMSE estimator, $\hat X(Y,Z) = \bE(X|Y,Z)$, has the following mean squared value:
\begin{align}
\label{e39}
\bE[\hat X^2(Y,Z)]&=\bE\left[\iint dx\, dx^\prime x\, x^\prime \,
P(x|Y,Z)\, P(x^\prime|Y,Z)\right]\nonumber\\
=\iint &dx\,dx^\prime x\, x^\prime 
\iint dy \, dz {P(x,y,z) P(x^\prime,y,z)
\over P(y,z)},
\end{align}
where the Bayes theorem was used to replace the {\it posterior} probabilities in terms of the joint 
PDFs. 

In terms of the integral,
\be
\label{e40}
K(z,y) \defeq \int dx \, x {P(x,y,z)\over \sqrt{P(z|y)}},
\ee
we may write the mean squared value of the joint MMSE estimator (\ref{e39}) as
\begin{align}
\label{e41}
\bE(\hat X^2(Y,Z)) &= \int {dy\over P(y)}
\int dz\, K^2(z,y)\cdot\int dz \, P(z|y)\nonumber\\  
&\geq \int{dy\over P(y)}\, \left|\int dz\, K(z,y)\sqrt{P(z|y)}\right|^2\nonumber\\
&=\int{dy\over P(y)}\, \iint dx\, dx^\prime x\, x\, P(x,y)\,P(x^\prime,y)\nonumber\\
&=\iint dx\, dx^\prime x\, x^\prime\, \int dy {P(x,y)P(x^\prime,y)\over P(y)}\nonumber\\
&=\bE_Y\{[\bE(X|Y)]^2\}.
\end{align}
The first equality follows from substituting the Bayes relation, $P(y,z) = P(y)\, P(z|y)$,
and definition (\ref{e40}) into expression (\ref{e39}) and from the unit normalization of the PDF $P(z|y)$;
the second line follows from the Cauchy-Schwarz inequality; the third line from a substitution of the definition
(\ref{e40}) and the identity, $\int dz\, P(x,y,z) = P(x,y)$; and the fourth line from 
an interchange of the order of the integrals.

Since the last expression in inequality (\ref{e41}) is simply the mean squared value of
the MMSE estimator, $\hat X(Y)$, relative to the measurement $Y$ alone, we
have arrived at the desired result,
\begin{align}
\label{e42}
\MMSE(Y,Z)&= \bE(X^2) - \bE[\hat X^2(Y,Z)]\nonumber\\
&\leq\bE(X^2)-\bE[\hat X^2(Y)] =\MMSE(Y),
\end{align}
where we have used the fact that $\bE[\hat X(Y)] = \bE[\hat X(Y,Z)]=E(X)$
to express the MSE, $\bE\{[\hat X(Y)-X]^2]\}$, as the difference of mean squared values of the prior and
the estimator.
Note that for the inequality (\ref{e42}) to hold, the two measurements are not required to be conditionally 
independent, given the input $X$.

The second property of MMSE relates to the case of multiple input parameters and how 
the error in the estimation of any one parameter is affected by the presence of the others. But
to fully appreciate this property, we must place it in the context of statistical information
processing to which I now turn.

\section{Role of Nuisance in Statistical Information Processing}
\label{MI+-}

It is well known that the fidelity of estimation of a parameter, defined here as the smallness of the
lower bound on the statistical variance of the estimator, decreases when other parameters are added to the
problem. These added parameters, when not of interest, are known as nuisance parameters,
and serve to reduce the fidelity, {\it i.e.,} increase the variance, of estimation of the parameter
of interest. The essence of this phenomenon is captured well by the FI matrix and its inverse whose
diagonal elements provide the Cram\'er-Rao lower bounds on the variances of an unbiased estimation of 
the parameters \cite{vanTrees68,Rao73}. 

A similar result must hold in the context of statistical information theory as well. It must be possible
to show that when the output variables $Y$ depend on two input parameters, $X$ and $U$,
that are distributed independently, then the MI between $X$ and $Y$ 
cannot be larger than the MI obtained by computing the MI between $X$ and $Y$ for a fixed value of $U$
first and then averaging it over the statistical distribution of the possible values of $U$.
The latter, averaged MI represents the information about $X$ successfully transmitted through the
information channel when $U$ is held fixed in each instance, so the statistical 
dispersion of $U$ does not corrupt the data relative to their capacity 
to carry information about $X$. I now prove this result.

Let us define $I^{(+)}(X;Y)$ as the MI in the case $U$ serves as a nuisance parameter, namely as
\begin{align}
\label{n15}
I^{(+)}(X;Y)&= I(X;Y)\nonumber\\
&= H(X)- H(X|Y),
\end{align}
where $H(X),H(X|Y)$ are defined as before. This expression for MI may also be written as the following
average over all three variables:
\begin{align}
\label{n16}
I^{(+)}(X;Y)&=- \bE \left\{\ln\left[{P(X,Y)\over P(X) P(Y)}\right]\right\} \nonumber\\
 &=-\int P(x,y,u) 
 \ln\left[{P(x,y)\over P(x) P(y)}\right] dx\, dy\, du,
\end{align}
as the integral over $u$ only 
affects the joint density $P(x,y,u)$, reducing it to the marginal, $P(x,y)$.

In the absence of nuisance, which is indicated by a $-$ superscript, 
the MI is the following $U-$averaged conditional MI:
\begin{equation}
\label{n17}
I^{(-)}(X;Y)=- \int P(x,y,u) 
 \ln\left[{P(x,y|u)\over P(x|u) P(y|u)}\right] dx\, dy\, du.
\end{equation}
Note that $I^{(-)}(X;Y)$ is the same as the more familiar conditional MI, $I(X;Y|U)$, so the difference between $I^{(+)}(X;Y)$
and $I^{(-)}(X;Y)$ is equivalently that between $I(X;Y)$ and its conditional version, $I(X;Y|U)$,
which, as is well known \cite{CoverThomas91}, can be of either sign.

In view of Jensen's inequality applied to the logarithm,
the difference between the two MIs, (\ref{n16}) and (\ref{n17}), has a lower bound,
\begin{align}
\label{n18}
I^{(-)}&(X;Y)-I^{(+)}(X;Y)\nonumber\\
&=-\int P(x,y,u) 
 \ln\left[{P(x,y|u)\over P(x|u) P(y|u)}\right] dx\, dy\, du\nonumber\\
&\geq-\ln\ \int P(x,y,u)  \left[{P(x,y)\, P(x|u)\, P(y|u)\over P(x,y|u)\, P(x)\, P(y)}\right] dx\, dy\, du\nonumber\\
&=-\ln\  \int  \left[{P(u)\, P(x,y)\, P(x|u)\, P(y|u)\over  P(x)\, P(y)}\right] dx\, dy\, du\nonumber\\
&=-\ln\ \int  \left[{P(x|u)\, P(y,u)\, P(x,y)\over P(x)\, P(y)}\right] dx\, dy\, du.
\end{align}
In obtaining the last two equalities above, I have used the Bayes rule twice, first via
the identity $P(x,y,u)=P(x,y|u)\, P(u)$, and then via the identity $P(y|u)\, P(u) = P(y,u)$.

When the variables $X$ and $U$ are statistically independent, $P(x|u)=P(x)$, the above inequality
simplifies greatly to the form
\begin{align}
\label{n19}
I^{(-)}&(X;Y)- I^{(+)}(X;Y)\nonumber\\
&\geq -\ln \ \iiint  \left[{ P(y,u)\, P(x,y)\over P(y)}\right] dx\, dy\, du\nonumber\\
 &= -\ln\,\iint P(x,y) dx\, dy\ = -\ln\ 1 = 0,
\end{align}
where I used the fact that $\int P(y,u)\, du= P(y)$ and the normalization of the joint PDF $P(x,y)$.
This proves our assertion. Note that since I have made no explicit use of the dimensionality 
of the input and output spaces in this proof, the result is valid for an arbitrary MIMO chennel.

\subsection{Analogous Result from Statistical Estimation Theory}

A correspondence may be drawn with analogous results from statistical estimation theory
using FI. One can consider two different estimation problems involving nuisance, one in which
the nuisance is also estimated and another in which it is not, which must be treated separately.

\paragraph{Estimation of Both Input and Nuisance Parameters}

The FI matrix relative to $X$ and $U$, when both are unknown, namely $\bJ$, may be expressed in terms
of the FI matrix relative to $X$, when $U$ is known, namely $\bJ_{XX}$, in the following block form:
\begin{equation}
\label{f1}
\bJ=\left[ \begin{array}{c | c}\bJ_{XX} & \bJ_{XU}\\
                               \hline
                                \bJ_{UX} & \bJ_{UU}
           \end{array}
    \right],
\end{equation}
where the matrix block $\bJ_{UU}$ refers to the FI matrix relative to $U$ alone, when $X$ is known,
and the off-diagonal blocks $\bJ_{XU}$ and $\bJ_{U X}$, which are transposes of each other, 
refer to the cross-sensitivity of the data likelihood relative to $X$ and $U$.
The presence of the cross-sensitivity matrices, $\bJ_{XU}$ and $\bJ_{U X}$, tends to increase
the CRBs since, as one may easily show \cite{Rao73} that, e.g., the $XX$ block of $\bJ^{-1}$ has the form
\begin{equation}
\label{f2}
\left(\bJ^{-1}\right)_{XX} = \left(\bJ_{XX} - \bJ_{XU}\bJ_{UU}^{-1}\bJ_{U X}\right)^{-1} \geq 
\left(\bJ_{XX}\right)^{-1},
\end{equation}
the matrix inequality following from the fact that $\bJ_{XU} \bJ_{U U}^{-1}\bJ_{U X}$ is a positive matrix. 
For the Bayesian case of priors on $X$ and $U$, assumed for the moment to be uncorrelated,
the FI matrices relative to these priors on $X$ and $U$ must be added to the
blocks $\bJ_{XX}$ and $\bJ_{UU}$, respectively, in expression (\ref{f1}).   
Adding these prior-information-based FI submatrices has, as expected, the opposite effect:
It decreases the CRBs on $X$ and $U$, thus improving the fidelity of estimation. 

\paragraph{Estimation of Input without Estimating Nuisance}

In this case, we must integrate over the statistical distribution of nuisance to obtain
the needed PDFs from their nuisance-free counterparts. We have, in particular, 
\begin{equation}
\label{fn1}
P(\bfy|\bfx) =\int P(\bfy|\bfx,\bfu)\, P(\bfu|\bfx)\, d\bfu,
\end{equation}
where the input, output, and nuisance parameters have been organized into three respective 
column vectors, $\bfX$, $\bfY$, and $\bfU$. If we take the nuisance and input variables
to be statistically uncorrelated, $P(\bfu|\bfx) = P(\bfu)$, then we may take
the gradient of Eq.~(\ref{fn1}) with respect to $\bfx$ simply,
\begin{equation}
\label{fn2}
\bnabla_x P(\bfy|\bfx) =\int \bnabla_x P(\bfy|\bfx,\bfu)\, P(\bfu)\, d\bfu.
\end{equation}
The inner product of this gradient vector with an arbitrary vector, $\blambda$, of the same length 
generates a scalar quantity 
\begin{equation}
\label{fn3}
\blambda^T \bnabla_x P(\bfy|\bfx) =\int \blambda^T\bnabla_x P(\bfy|\bfx,\bfu)\, P(\bfu)\, d\bfu.
\end{equation}
Upon writing the integrand in Eq.~(\ref{fn3}) as the bilinear product 
\begin{align}
\label{fn4}
[P^{1/2}&(\bfu) P^{-1/2}(\bfy|\bfx,\bfu)\blambda^T\bnabla_x P(\bfy|\bfx,\bfu)] \nonumber\\
&\times [P^{1/2}(\bfu) P^{1/2}(\bfy|\bfx,\bfu)],
\end{align}
squaring both sides of that equation, and then using the Cauchy-Schwarz inequality, we arrive
at the inequality
\begin{align}
\label{fn5}
[\blambda^T \bnabla_x P(\bfy|\bfx)]^2 &\leq \int d\bfu\, P(\bfu) {1\over P(\bfy|\bfx,\bfu)}[\blambda^T
\bnabla_x P(\bfy|\bfx,\bfu)]^2\nonumber\\
&\qquad\times \int P(\bfu)\, P(\bfy|\bfx,\bfu)\, d\bfu. 
\end{align}
Since the last $\bfu$-integral above evaluates simply to $P(\bfy|\bfx)$ according to Bayes rule,
by dividing both sides by $P(\bfy|\bfx)$, integrating over $d\bfy$, and finally
averaging over $\bfX$, we obtain the desired inequality, 
\begin{align}
\label{fn6}
\blambda^T \bJ^{(+)}(\bfY|\bfX)\, \blambda &\leq \blambda^T\int d\bfu\, P(\bfu) \bJ_\bfu(\bfY|\bfX)\,\blambda
\nonumber\\
&=\blambda^T\bJ^{(-)}(\bfY|\bfX)\, \blambda,
\end{align}
where the FI matrices in the presence and absence of nuisance are defined as 
\begin{align}
\label{fn7}
\bJ^{(+)}(\bfY|\bfX)&\defeq \iint d\bfx\, d\bfy \, P(\bfx) \, P(\bfy|\bfx)
\left[{1\over P(\bfy|\bfx)}\right]^2 \nonumber\\
&\qquad \times\bnabla_x P(\bfy|\bfx)\, \bnabla_x^T P(\bfy|\bfx)  \nonumber\\
\bJ^{(-)}(\bfY|\bfX)&\defeq \int d\bfu\, P(\bfu) \iint d\bfx\, d\bfy \, P(\bfx) \, P(\bfy|\bfx,\bfu)
\nonumber\\
&\times \left[{1\over P(\bfy|\bfx,\bfu)}\right]^2 
\bnabla_x P(\bfy|\bfx,\bfu)\, \bnabla_x^T P(\bfy|\bfx,\bfu).  
\end{align}
Note that for statistically mutually independent input and nuisance parameters, 
the prior-based FI for the input paramaters is the same whether the nuisance
parameters are present or absent. It then follows from the 
the non-negative-definiteness of the difference of the data-based FIs,
$\bJ^{(-)}(\bfY|\bfX)-\bJ^{(+)}(\bfY|\bfX)$, implied by 
relation (\ref{fn6}), that the corresponding difference between the 
sums of data-based and prior-based FIs is also non-negative-definite. 
This result embodies the fact
that in general nuisance parameters even when they are not estimated, if statistically independent
of the input parameters of interest, degrade the fidelity with which the latter can be estimated \cite{DMR94}.

\subsection{Statistical Correlation of $X$ and $U$ Priors}

For the more general case when $X$ and $U$ are statistically correlated, $U$ may indeed carry 
information about $X$ through their correlation, in which case the RHS of the inequality (\ref{n18})
may be negative allowing for $I^{(+)}$ to exceed $I^{(-)}$.
As I noted before, in this general case the inequality (\ref{n18}) can be of either sign.

The corresponding result from statistical estimation theory is based on the fact that any information
that $U$ has about $X$ through its correlations with it yields an additional FI submatrix, $\bJ^{(U)}_{XX}$,
to be added to the $\bJ_{XX}$ block in Eq.~(\ref{f1}). This submatrix represents information that $U$
carries about $X$ through the first-order sensitivity of $P(u|x)$ on $x$. Unlike the coupling of $U$ 
to the data alone, such additional prior information can reduce the CRBs on estimating $X$.
When the nuisance parameters are {\em not} estimated, but are correlated with the input
parameters of interest, the basic relation (\ref{fn2})
used to obtain the desired inequality (\ref{fn6}) is itself not valid.
Also, the prior-based FIs are not necessarily the same with and without nuisance. As a result,
the data-based FI or prior-based FI or both may contain more information about the
parameters of interest in the presence of nuisance than in its absence. 

\subsection{Two Illustrative Examples}

As a first example, let us consider a Gaussian additive channel with additive noise $N$,
\begin{equation}
\label{n21a}
Y = aX+bU+N,
\end{equation}
where $X$, $U$, and $N$ are all independently normally distributed as follows: 
\begin{equation}
\label{n21b}
X \sim {\cal N} (\bar X,\sigma_X^2),\ \ 
U \sim {\cal N} (\bar{U},\sigma_{U}^2),\ \ 
N \sim {\cal N} (0,\sigma_N^2).
\end{equation}
Thus the marginal PDF for $Y$ as well as its various conditional PDFs are all 
Gaussian too,
\begin{align}
\label{n22}
Y &\sim {\cal N} (a\bar X + b\bar{U},a^2\sigma_X^2+b^2\sigma_{U}^2+\sigma_N^2),\nonumber\\ 
Y|X &\sim {\cal N} (a X + b\bar{U},b^2\sigma_{U}^2+\sigma_N^2),\nonumber \\ 
Y|{U} &\sim {\cal N} (a\bar X + b{U},a^2\sigma_X^2+\sigma_N^2),\nonumber \\ 
Y|X,{U} &\sim {\cal N} (a X + b{U},\sigma_N^2). 
\end{align}

In view of the variances given in relations (\ref{n22}), we may easily write down the
MIs of interest here using the well known expression for the differential entropy for a 
Gaussian additive channel \cite{CoverThomas91},
\begin{align}
\label{n23}
I^{(+)}(X;Y) &= I(X;Y) = {1\over 2} \ln \left(1+{a^2\sigma_X^2\over b^2\sigma_{U}^2+\sigma_N^2}\right); 
\nonumber\\
I^{(-)}(X;Y) &= I(X;Y|{U}) = {1\over 2} \ln \left(1+{a^2\sigma_X^2\over \sigma_N^2}\right). 
\end{align}
Since $b^2\sigma_{U}^2 > 0$, it follows that $I^{(-)}(X;Y) > I^{(+)}(X;Y)$.
For uncorrelated $X$ and ${U}$, the latter serves as a nuisance relative to information about the former
in the sense that the terms $b{U}$ and $N$ in the model (\ref{n21a})
simply combine to yield an increased noise variance, $b^2\sigma_{U}^2+\sigma_N^2$,
on the determination of $X$ from $Y$.
But when the nuisance is removed by holding ${U}$ fixed in each measurement, 
the noise variance is lower at $\sigma_N^2$, leading to increased information about $X$.

As my second example, let us modify the Gaussian channel represented by Eqs.~(\ref{n21a}) and (\ref{n21b})
simply to include a correlation between $X$ and ${U}$, so only the first of the relations in Eq.~(\ref{n21b})
is changed to the conditional relation
\begin{equation}
\label{n24}
X|U \sim {\cal N}(\alpha U,\sigma_{X|U}^2),
\end{equation}
while the remaining relations are unchanged.
In the limit that $\alpha\to 0$, the variables $X$ and $U$ become uncorrelated as in the previous example.
Thus the largeness of $|\alpha|\sigma_{U}$ in relation to $\sigma_{X|{U}}$
may be regarded as the strength of the correlation between $X$ and ${U}$.

In view of the relation (\ref{n24}) and the fact that $U\sim {\cal N} (\bar{U},\sigma_{U}^2)$, 
the marginal PDF $P(x)$ is also Gaussian. 
The conditional PDF, $P(y|u)$, for $Y$, given ${U}$, may be computed by integrating $P(y|x,u) P(x|u)$
over $x$. The marginal PDF $P(y)$ is then obtained by integrating $P(y|u)P(u)$ over $u$.
Using standard analysis involving Gaussian integrals, we may derive
the following marginal and conditional PDFs:
\begin{align}
\label{n25}
X&\sim {\cal N}(\alpha\bar{U},\sigma_X^2\equiv\sigma_{X|U}^2 +\alpha^2 \sigma_{U}^2);\nonumber\\
Y|{U}&\sim {\cal N}((a\alpha+b){U},\sigma_N^2+a^2\sigma_{X|{U}}^2);\nonumber\\
Y&\sim {\cal N}((a\alpha+b)\bar{U},\sigma_Y^2),
\end{align}
where the $Y$-variance may be expressed as
\begin{equation} 
\label{n25a}
\sigma_Y^2=\sigma_N^2+a^2\sigma_{X|{U}}^2+(a\alpha+b)^2\sigma_{U}^2.
\end{equation}
Having evaluated $P(x)$, we may now evaluate $P(u|x)$, via Bayes rule, as the ratio $P(u)P(x|u)/P(x)$,
\begin{align}
\label{n26}
P(u|x) &= {\sqrt{\sigma_{X|{U}}^2 +\sigma_{U}^2}\over \sqrt{2\pi}\sigma_{U}\sigma_{X|{U}}}
\exp\left[-{(u-\bar{U})^2\over 2\sigma_{U}^2}-
{(x-\alpha u)^2\over 2\sigma_{X|{U}}^2}\right.\nonumber\\
&\qquad+\left.{(x-\alpha \bar{U})^2\over 2(\sigma_{X|{U}}^2+\alpha^2\sigma_{U}^2)}\right]\nonumber\\
&={1\over \sqrt{2\pi\sigma_{U|X}^2}}\exp\left[-{\left(u-\bar{U}_{|x}
\right)^2\over 2\sigma_{U|X}^2}\right],
\end{align}
where the conditional mean, $\bar{U}_{|x}$, and variance, $\sigma_{U|X}^2$, are given by the expressions
\begin{align}
\label{n27}
\bar{U}_{|x}&=\sigma_{{U}|X}^2\left({\alpha x \over\sigma_{X|{U}}^2}+ {\bar{U}\over
\sigma_{U}^2}\right);\nonumber\\
{1\over \sigma_{{U}|X}^2}&={1\over \sigma_{U}^2}+{\alpha^2\over \sigma_{X|{U}}^2}.
\end{align}

By multiplying $P(u|x)$ with $P(y|u,x)$ and integrating over $u$, 
we may obtain the last of the needed conditional PDFs, namely $P(y|x)$, 
\begin{equation}
\label{n28}
P(y|x) = {1\over \sqrt{2\pi\sigma_{Y|X}^2}}
\exp\left[-{(y-ax-b\bar{U}_{|x})^2\over 2\sigma_{Y|X}^2}\right],
\end{equation}
where the conditional mean and variance may be expressed as
\begin{align}
\label{n29}
\bE(Y|x)&= \left(a+{\alpha b\sigma_U^2\over
\sigma_X^2}\right)x+{b\bar U\sigma_{X|U}^2\over\sigma_{X|U}^2+\alpha^2\sigma_U^2};
\nonumber\\ 
\sigma_{Y|X}^2 &= \sigma_N^2+b^2{\sigma_U^2\sigma_{X|U}^2
\over\sigma_{X|U}^2+\alpha^2\sigma_U^2}.
\end{align}

We are now in a position to write down both $I^{(+)}(X;Y)$ and $I^{(-)}(X:Y)$ by use of the Gaussian-channel
entropy formula in terms of the variances of $P(y)$, $P(y|x)$, $P(y|u)$, and $P(y|x,u)$,
\begin{align}
\label{n30}
I^{(+)}(X;Y) &= I(X;Y)\nonumber\\
&= {1\over 2} \ln \left[{\sigma_N^2+a^2\sigma_{X|{U}}^2+ (b+a\alpha)^2\sigma_{U}^2
\over \sigma_N^2+b^2\sigma_{{U}|X}^2}\right]; \nonumber\\
I^{(-)}(X;Y) &= I(X;Y|U)\nonumber\\
&= {1\over 2} \ln \left(1+{a^2\sigma_{X|{U}}^2\over \sigma_N^2}\right). 
\end{align}
Note that in the limit of $\sigma_{X|{U}}\to 0$, the two variables, $X$ and ${U}$, are infinitely tightly
coupled. In effect, $X=\alpha {U}$, and the data $Y$ carry no information about $X$ when ${U}$ is held fixed. 
This is seen in the 
relation (\ref{n30}). By contrast, $I^{(+)}(X;Y)$ is finite in this limit, and thus trivially exceeds
$I^{(-)}(X;Y)$.
The other limit, $\alpha\to 0$, returns us to the case of uncorrelated $X$ and $U$ variables for which the results
(\ref{n23}) are recouped and $I^{(-)}(X;Y)$ exceeds $I^{(+)}(X;Y)$.
The more general cases in which neither of these limits is a good approximation are illustrated in Fig.~3.

\begin{figure*}
\centerline{
\subfigure{\includegraphics[width=6in]{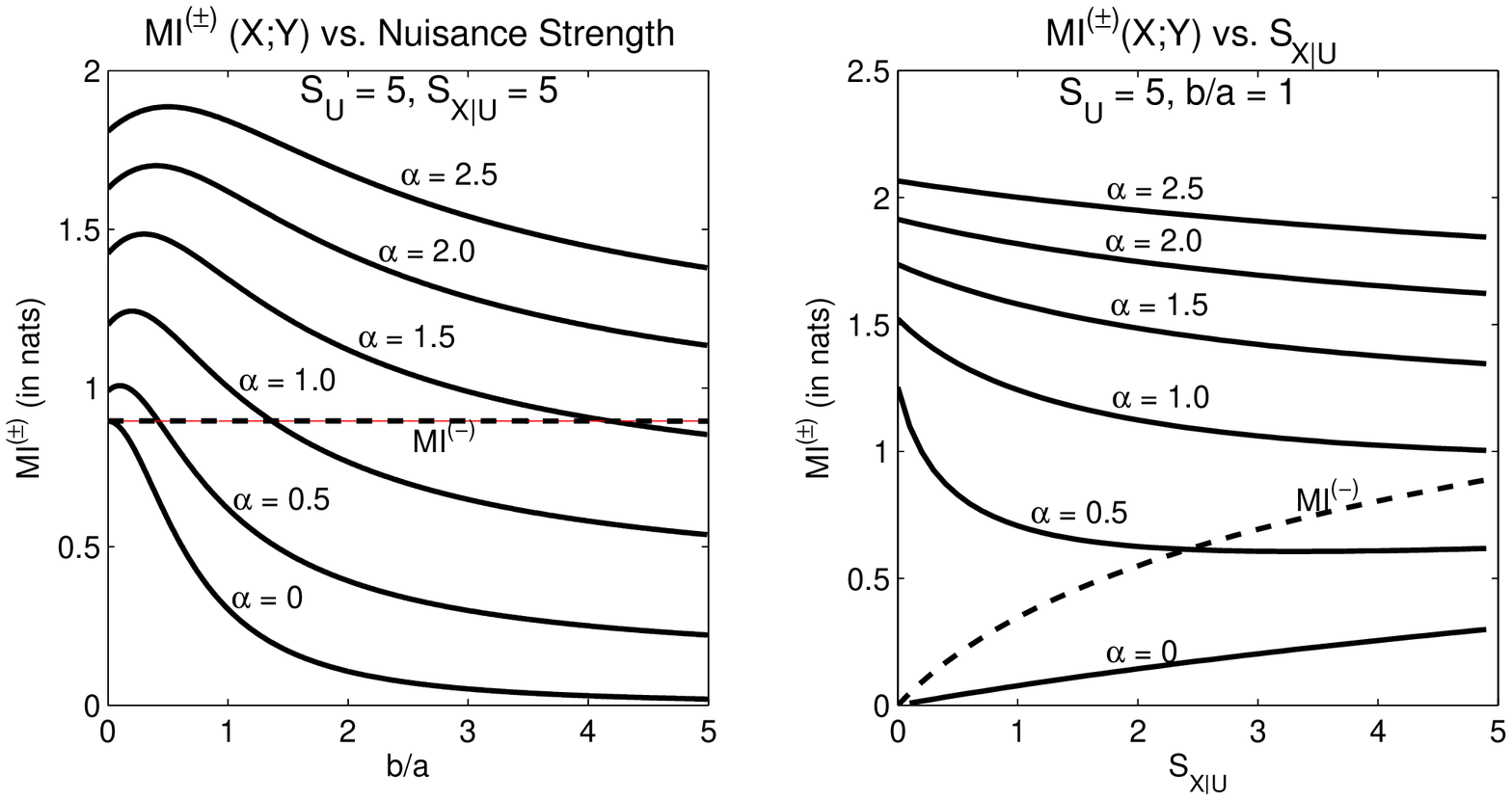}}}
\vspace{1cm}
\centerline{\subfigure
{\includegraphics[width=6in]{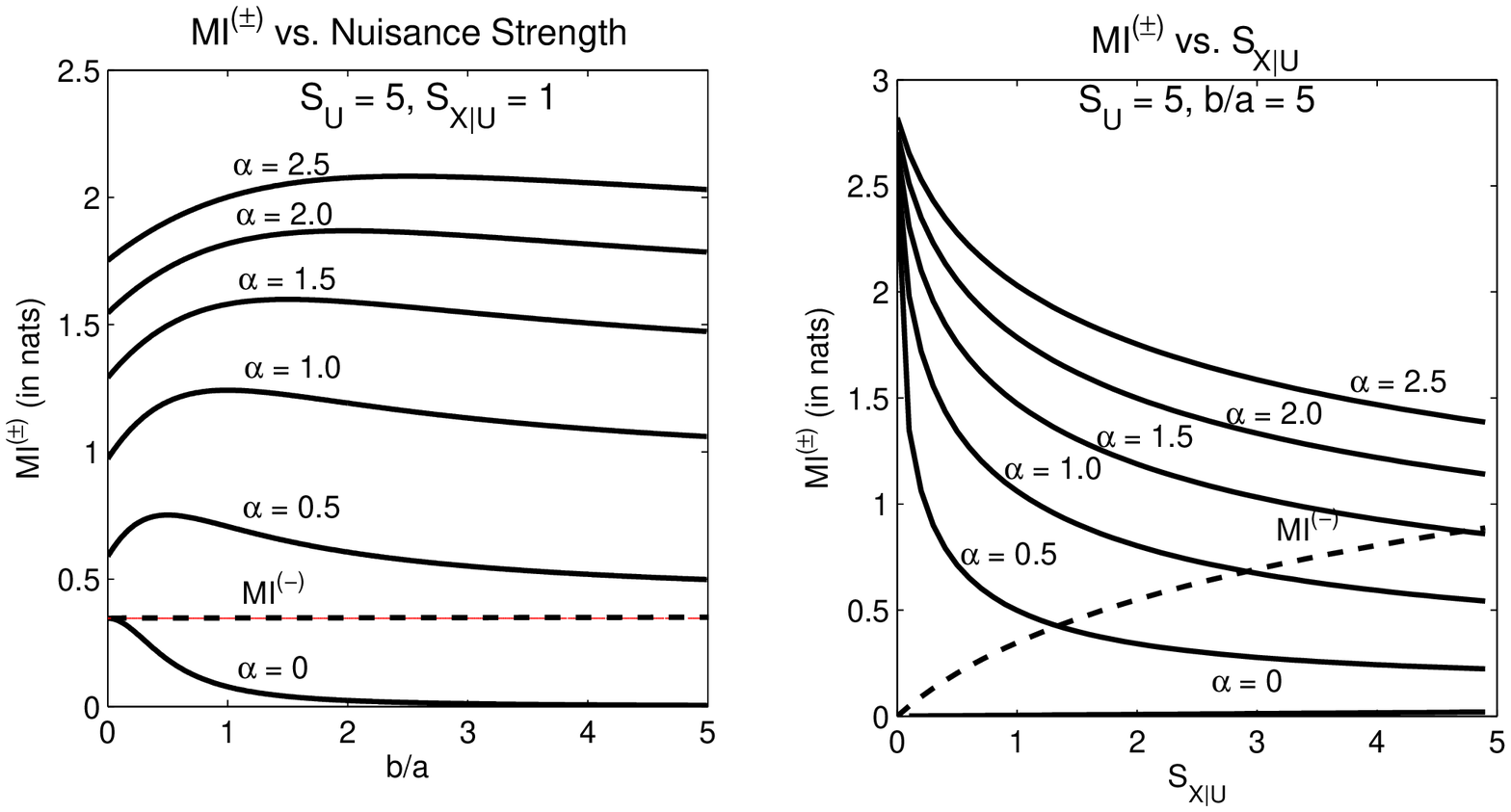}}}
\caption{\label{fig:f3} Plots of $I^{(\pm)}(X;Y)$ vs relative strength, $b/a$, of the nuisance parameter (left top
and bottom panels) and vs. weakness of coupling between $X$ and ${U}$, as measured by $a^2\sigma_{X|{U}}^2/\sigma_N^2$ 
(right top and bottom panels). The bottom panels refer to a tighter $X-{U}$ coupling (left panel) and larger nuisance 
parameter strength (right panel) than the corresponding figures in the top panels.}
\end{figure*}
I plot here $I^{(+)}(X;Y)$ (solid curves) and $I^{(-)}(X;Y)$ (dashed curves) for the case of normalized ${U}$ variance,
$S_U= a^2 \sigma_{U}^2/\sigma_N^2$, equal to 5. Each variance is normalized the same way by multiplying it
with $a^2/\sigma_N^2$. Six different values of $\alpha$, which determines the largness of
the conditional mean of $X$, given ${U}$, were used to generate the various $I^{(+)}$ curves.
By contrast, $I^{(-)}$ is independent of $\alpha$, as seen from the single dashed curve on each plot.
A number of observations can be made from these plots. First, for $\alpha=0$, the variables $X$ and ${U}$
are uncorrelated, so in this case the plot of MI in the presence of the nuisance variable, ${U}$, lies below
that for MI when uninfluenced by the nuisance variable.
Second, as $\alpha$ increases, the coupling of $X$ and ${U}$ becomes increasingly less sensitive
to the noise in ${U}$. This means that when $Y$ is measured, its value reveals more information about $X$
than when $\alpha$ is smaller. 
When the nuisance is removed, {\it i.e.,} ${U}$ is held fixed,
then a change of $\alpha$ merely changes the mean value of $X$, leaving its variance unchanged, which
is the reason why $I^{(-)}$ depends neither on $\alpha$ nor on $\sigma_U$. 
Third, as the relative strength, $b/a$, of
the nuisance increases while $\alpha$ is held fixed, $I^{(+)}$ increases initially since the data $Y$
possess an increasing amount of information about $X$ through the latter's coupling to ${U}$. However, increasing
the strength of the $b{U}$ term in Eq.~(\ref{n21a}) to large values leads to the data becoming more corrupted
than helped by the nuisance, which leads to an eventual decrease of the MI.
These two competing tendencies lead to a maximum for each curve (left top and bottom), with 
the location of the maxima shifting to larger nuisance-parameter strength values with increasing $\alpha$.
Fourth, comparing the plots for the smaller vs. larger values of $S_{X|U}$ (left panels),
we see that the tighter the $X$-$U$ coupling the softer the degradation of MI$^{(+)}$
with increasing strength of the nuisance parameter. 
Finally, as seen from the right-hand panels of the figure, an infinitely tight coupling between $X$ and ${U}$ (for $S_{X|{U}}=0$)
yields, through the sensitivity of data to ${U}$, information about $X$ as well. This information about
$X$ degrades when $S_{X|{U}}$ increases to finite values, the more so the smaller the parameter $\alpha$.

\section{MMSE in the Presence of Nuisance Parameters}

When multiple input parameters must all be estimated from the 
same measurement(s), one expects the MMSE, like the CRB, for estimating any of the parameters 
to be higher than if the others were not present. I prove this result next.

Let $X,U$ be two input parameters to be estimated from data $Y$. 
Let $P(x,u)$ be the joint prior on the inputs. The MMSE estimator for $X$ in the absence of the nuisance $U$
can be defined in terms of the conditional MMSE estimator,
\begin{equation}
\label{n31}
\hat X_U (Y)= \int x \, P(x|Y,U=u)\, dx,
\end{equation}
given $U=u$.
It is the MMSE estimator of $X$ for a given value of $U$. Its mean squared value has an expression analogous to
that found in (\ref{e41}), 
\begin{align}
\label{n32}
\bE(\hat X_u^2(Y))&= \int du\, P(u) \iint dx\, dx^\prime x\, x^\prime P(x|u)\, P(x^\prime|u)
\nonumber\\
&\times\int dy\, {P(y|x,u)\, P(y|x^\prime,u)\over P(y|u)}\nonumber\\
&=\int {dy\over P(y)}\, \int du \, K^2(y,u)\cdot\int du\, P(y|u)\, P(u),
\end{align}
where $K(y,u)$ stands for the function
\begin{equation}
\label{n33}
K(y,u)\defeq \sqrt{P(u)\over P(y|u)}\, \int \, x\, P(y|x,u)\, P(x|u)\, dx.
\end{equation}
We also used the Bayes-rule identity, $\int P(y|u)\, P(u)\, du = P(y)$,
to arrive at the last line of Eq.~(\ref{n32}).

A use of the Cauchy-Schwarz inequality in Eq.~(\ref{n32}) shows that the mean squared value of the MMSE estimator
in the absence of nuisance has the lower bound
\begin{align}
\label{n34}
\bE(\hat X_u^2(Y)) &\geq  
\int {dy\over P(y)}\, \left|\int du\, K(y,u) \, \sqrt{P(y|u)\, P(u)}\right|^2 \nonumber\\
&=\int {dy\over P(y)}\,\left|\iint dx\,  du\, x\, x^\prime P(x|u)\, P(y|x,u)\, P(u)\right|^2 \nonumber\\ 
&=\int {dy\over P(y)}\,\left|\int dx\, x \, P(x,y)\right|^2 \nonumber\\ 
&= \bE[\hat X^2(Y)],
\end{align}
where a simple substitution of $K(y,u)$ from Eq.~(\ref{n33}) was used to obtain the second relation,
Bayes rule to obtain the third relation, and the definition of the MMSE estimator $\hat X(Y)$
as the posterior mean of $X$, namely $\int dx\, x P(x,y)/P(y)$, to arrive at the final relation.
Since the MMSE, as we have noted earlier, may be expressed simply as
the mean squared value of $X$ minus the mean squared value of the MMSE estimator, 
the desired inequality between the MMSE without and with nuisance follows immediately,
\begin{equation}
\label{n35}
\MMSE^{(-)}(X) \leq \MMSE^{(+)}(X).
\end{equation}

We deduce from this important result that the presence of the nuisance parameter can {\it never} lower the
MMSE below that obtained in its absence, {\it i.e.,} when the nuisance has a known value, regardless of whether the priors on $X$ and 
the nuisance $U$ is statistically correlated or not. This seems to exclude the possibility 
that $U$ if suitably correlated with $X$ may serve, as we observed in Sec.~\ref{MI+-} in the context of MI, 
as a source of additional information for $X$. The answer to this apparent paradox 
may be found in the way MMSE is defined. Since given a value of the nuisance $u$, the MMSE estimator minimizes the 
MSE relative to the corresponding conditional prior, $P(x|u)$, on $X$ and measurement PDF $P(y|x,u)$,
the nuisance-averaged MMSE is not characterizable as the MSE for a single, nuisance-averaged MMSE estimator.
The MMSE metric thus may not possess the same degree of specificity as the MI or FI metrics when
the effect of nuisance must be quantified.

\subsection{Gaussian Channel and Gaussian Prior}

We now illustrate the effect of nuisance on the MMSE with our previous example of a Gaussian channel 
for which some of the relevant PDFs are given in Eqs.~(\ref{n24})-(\ref{n28}). 
What we need are the MMSE$^{(\mp)}$ estimators, namely $\hat X_u(Y)$ given by expression (\ref{n31}),
and $\hat X(Y)$ by (\ref{e15}). As is well known from the theory of MMSE \cite{Kay93} for Gaussian
priors and Gaussian channel PDFs, each MMSE estimator
may be expressed as the inverse-variance-weighted sum of its prior and measurement based estimates,
\begin{align}
\label{n36}
\hat X_u(Y)&=f^{(-)}\left({Y-bu\over a\sigma_N^2}+{\alpha u\over a^2\sigma_{X|U}^2}\right); \nonumber\\ 
\hat X(Y)&=f^{(+)}\Biggl[{\left(Y-b\bar U{\sigma_{X|U}^2\over
\sigma_X^2}\right)
\left(a+{b\alpha\sigma_U^2\over\, \sigma_X^2}\right)
\over \sigma_{Y|X}^2}\nonumber\\
&\qquad\qquad + {\alpha \bar U\over \sigma_X^2}
\Biggr],
\end{align}
where the multipiers $f^{(\mp)}$ are given by
\begin{align}
\label{n37}
{1\over f^{(-)}}&={1\over \sigma_N^2}+{1\over a^2\sigma_{X|U}^2}; \nonumber\\ 
{1\over f^{(+)}}&={\left(a+{b\alpha \sigma_U^2\over\sigma_X^2}\right)^2\over 
\sigma_{Y|X}^2}+{1\over \sigma_X^2}
\end{align}
and the unconditional $X$-variance, $\sigma_X^2$, may be expressed as
\begin{equation}
\label{n38}
\sigma_X^2 =\sigma_{X|U}^2+\alpha^2\sigma_U^2.
\end{equation}
The various data and prior based estimates and variances
used in arriving at the expressions (\ref{n36}) have been inferred
from the mean values and variances of the PDFs given in Eqs.~(\ref{n24})-(\ref{n29}).

\paragraph{MMSE in the Absence of Nuisance}

To compute the mean squared values of these estimators, we first subtract and add the
appropriate mean values of $Y$ from it in the expressions (\ref{n36}) and then use the
fact that $\bE[(\delta Y+q)^2]=\bE[(\delta Y)^2]+q^2$, 
where $\delta Y$ is the deviation of $Y$ from its mean and $q$ is any quantity independent of $Y$.
For the MMSE$^{(-)}$ estimator, the mean we subtract and add is the conditional mean of $Y$, given $u$,
namely $(\alpha a+b)u$, so the following conditional squared mean value for it, given $u$, results:
\begin{align}
\label{n39}
\bE[\hat X_u^2(Y)|u]&=f^{(-)2}\left[{a^2\sigma_{X|U}^2+\sigma_N^2\over a^2\sigma_N^4}
+\alpha^2u^2/f^{(-)2}\right]\nonumber\\
&={\sigma_{X|U}^2/\sigma_N^2\over {1\over\sigma_N^2}+{1\over a^2\sigma_{X|U}^2}}
+\alpha^2u^2.
\end{align}
An averaging of this expression over $u$ with the help of the result
$\bE(U^2)=\bar U^2+\sigma_U^2$ then yields the required mean squared value of the
MMSE$^{(-)}$ estimator. Subtracting this squared mean value from $\bE(X^2)$,
the latter being simply $\alpha^2\bar U^2+\sigma_X^2$, generates, according to Eq.~(\ref{e16}),
the MMSE$^{(-)}$,
\begin{align}
\label{n40}
\MMSE^{(-)}&=\sigma_X^2-\alpha^2\bar U^2-{\sigma_{X|U}^2/\sigma_N^2\over 
{1\over\sigma_N^2}+{1\over a^2\sigma_{X|U}^2}}\nonumber\\
&={\sigma_{X|U}^2\sigma_N^2\over a^2\sigma_{X|U}^2+\sigma_N^2},
\end{align}
where use was made of relation (\ref{n38}) in the second line.

\paragraph{MMSE in the Presence of Nuisance}

Subtracting and adding the mean value of $Y$, namely $(a\alpha+b)\bar U$, 
from $Y$ inside the expression (\ref{n36}) for the estimator $\hat X(Y)$
and the squaring and averaging over $Y$ generates the following mean squared
value of the MMSE estimator in the presence of nuisance:
\begin{align}
\label{n41}
\bE[\hat X^2(Y)]&=f^{(+)2}{\sigma_Y^2\over \sigma_{Y|X}^4}
\left[{a\sigma_X^2+\alpha b\sigma_U^2\over\sigma_X^2}\right]^2\nonumber\\
&+f^{(+)2}{\alpha^2\bar U^2\over \sigma_X^4}
\left[{(a\sigma_X^2+\alpha b\sigma_U^2)^2\over\sigma_{Y|X}^2\sigma_X^2}+1\right]^2\nonumber\\
&=\sigma_Y^2 \left[{(a\sigma_X^2+\alpha b\sigma_U^2)^2\over\sigma_{Y|X}^2
+(a\sigma_X^2+\alpha b\sigma_U^2)^2/\sigma_X^2}\right]^2\nonumber\\
&+\alpha^2\bar U^2,
\end{align}
where use was made of the definition (\ref{n37}) of $f^{(+)}$ to simplify both terms on the
RHS. In view of relations (\ref{n29}) for the conditional mean and variance of $Y$, given $X=x$, and
the fact that all PDFs are Gaussian, we may express the unconditional variance of $Y$, namely $\sigma_Y^2$,
as the sum of conditional variance, $\sigma_{Y|X}^2$, given $X$, and $(a+\alpha b*\sigma_U^2/\sigma_X^2)^2$
times $\sigma_X^2$. This observation greatly simplifies the preceding expression,
\begin{equation}
\label{n42}
\bE[\hat X^2(Y)]=\alpha^2\bar U^2+ {(a\sigma_X^2+\alpha b\sigma_U^2)^2\over\sigma_Y^2}.
\end{equation}

The MMSE now follows from subtracting expression (\ref{n42}) from $\bE(X^2)= \alpha^2\bar U^2+\sigma_X^2$, a result
that can be simplified further in view of the relation between $\sigma_Y^2$ and $\sigma_{Y|X}^2$ 
that we just noted in the previous paragraph, 
\begin{equation}
\label{n43}
\MMSE^{(+)}={\sigma_{Y|X}^2\sigma_X^2\over\sigma_Y^2}.
\end{equation}
By using relations (\ref{n29}), (\ref{n27}), the alternate form of $\sigma_Y^2$
given by relation (\ref{n25}), and $\sigma_X^2=\sigma_{X|U}^2+\sigma_U^2$, 
we may express the MMSE in the presence of nuisance in the more explicit form
\begin{equation}
\label{n44}
\MMSE^{(+)}={\sigma_N^2\sigma_{X|U}^2+\alpha^2\sigma_N^2\sigma_U^2+b^2\sigma_{X|U}^2\sigma_U^2
\over \sigma_N^2+a^2\sigma_{X|U}^2+(a+\alpha b)^2\sigma_U^2}.
\end{equation}

In this form, we may easily compare MMSE$^{(+)}$ to the corresponding result (\ref{n40})
for MMSE in the absence of nuisance.
A sequence of steps involving simple algebraic manipulations, followed by a use
of the inequality, $f^2+g^2\geq 2fg$, easily confirms the general result proved earlier that
the presence of nuisance parameters can never reduce the MMSE for the estimation of the parameter of interest,
\begin{equation}
\label{n45}
\MMSE^{(+)}-\MMSE^{(-)}\geq 0.
\end{equation}

This result is illustrated in Fig.~4 where we plot both MMSE$^{(\pm)}$, in units of $\sigma_N^2/a^2$, as 
functions of the 
variable $\chi = a^2\sigma_{X|U}^2/\sigma_N^2$ for different values of the nuisance coupling parameter, $\eta= \alpha a/b$.
The reciprocal of $\chi$ is a measure of the strength of the statistical correlation between $X$ and nuisance $U$,
while $\eta$ represents the ability of nuisance to carry information about $X$ through its statistical correlations with $X$.
As $\chi$ becomes larger, the prior on $X$ becomes broader and the measurement becomes increasingly more dominant
in controlling the MMSE whether the nuisance is absent or present. But, as expected, MMSE$^{(-)}$ does 
not depend on the coupling parameter $\eta$ or the nuisance-parameter SNR defined as SNR$_U= b^2\sigma_U^2/\sigma_N^2$.
On the other hand, MMSE$^{(+)}$ decreases with increasing $\eta$ since the nuisance becomes increasingly more
effective - and the data $Y$ increasingly less so - in controlling the MSE. With increasing SNR$_U$, from 10 to 100
between the two panels of the figure, the nuisance causes an increased error in estimating $X$, as its increased 
variance leads to an increased variance of the prior on $X$. But in no event does the MMSE in the presence of nuisance 
fall below the MMSE without nuisance. The optimal condition under which the presence of nuisance does not degrade the MMSE,
{\em i.e.,} MMSE$^{(+)}$ = MMSE$^{(-)}$, is achieved when $\chi=\eta$, as seen from the figures and can also be easily shown analytically
from the expressions (\ref{n40}) and (\ref{n44}).
\begin{figure*}
\subfigure{\includegraphics[width=3.5in]{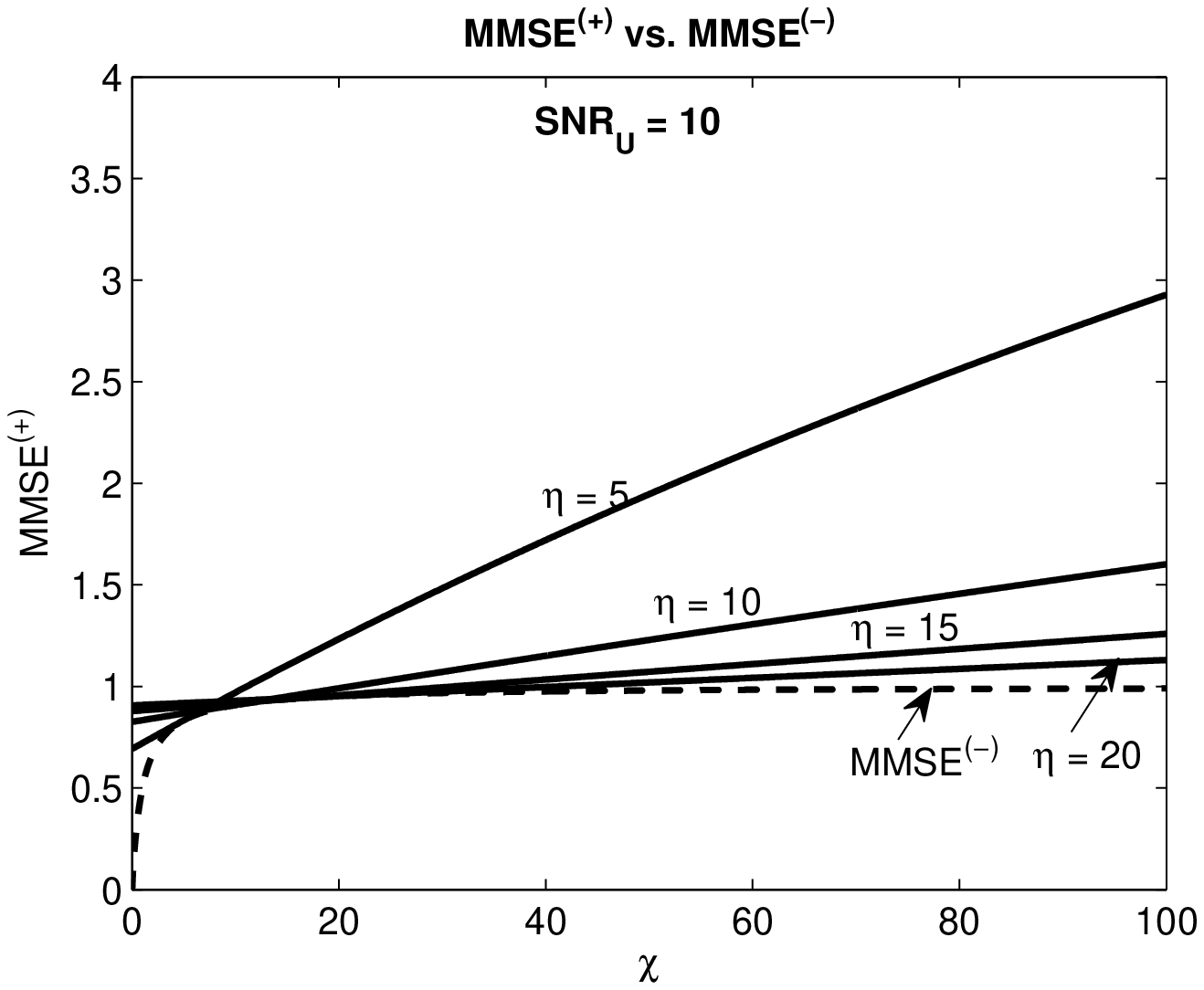}}
\subfigure
{\includegraphics[width=3.5in]{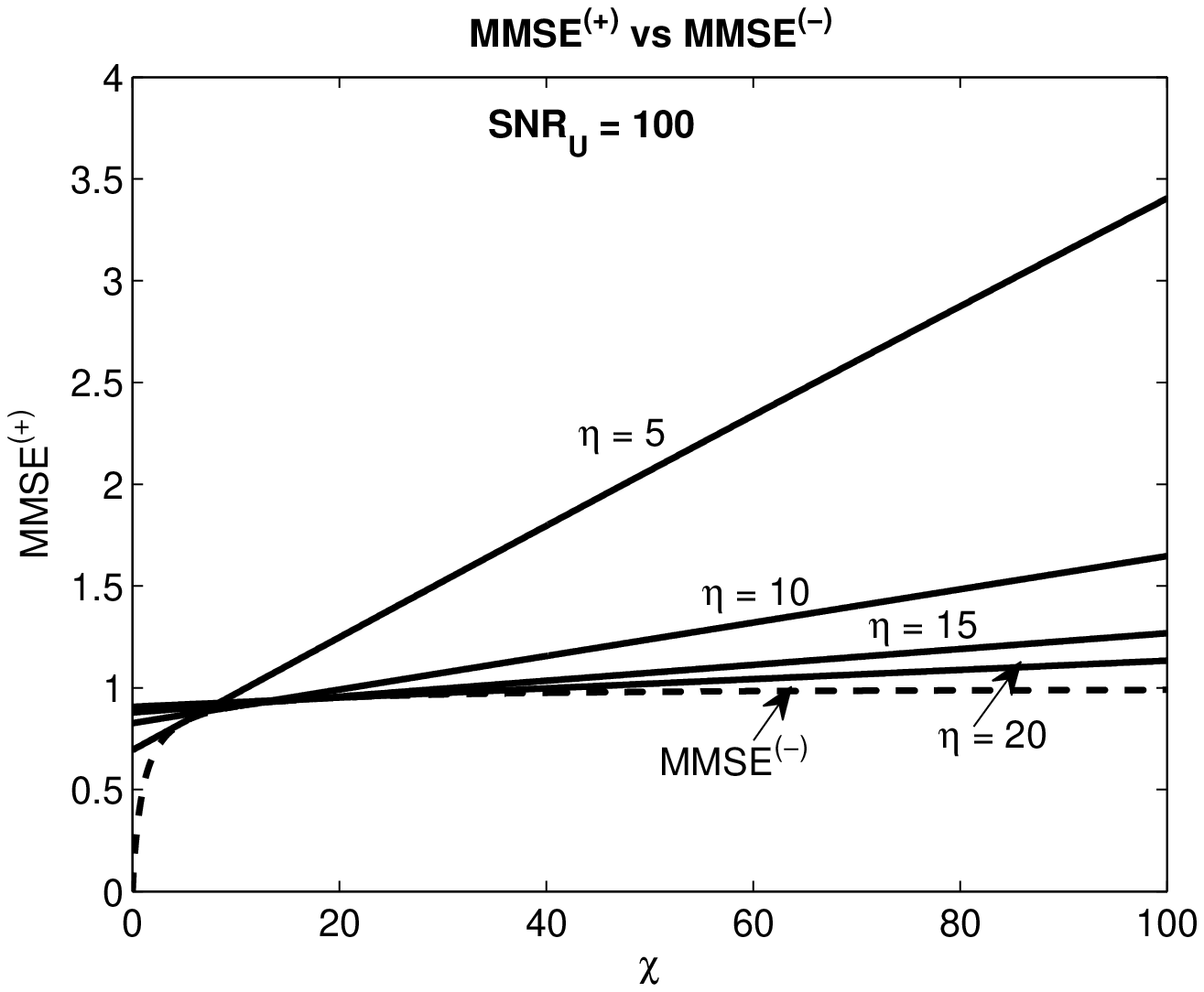}}
\caption{\label{fig:f4} Plots of MMSE vs. $\chi=a^2\sigma_{X|U}^2/\sigma_N^2$
in the absence of nuisance (dotted curve) and in the presence of nuisance (solid curves),
for four different values of $\eta=\alpha a/b$, the nuisance coupling parameter. The left and right panels of the figure are
for the values 10 and 100 of the nuisance-parameter SNR, ${\rm SNR}_U\equiv b^2\sigma_U^2/\sigma_N^2$, respectively.}
\end{figure*}

\section{Conclusions}

In this paper I have derived a number of previously unknown 
relationships between mutual information and the minimum error of estimating 
a parameter from its measurements. A seoond order linear relation between MI and 
a prior-averaged, squared-deviation-weighted form of the FI
accords added significance to the phrase ``information" when describing the latter
even though its chief claim to this phrase has been in the sense
of being the reciprocal of estimation error.

A second, more important relation between information and estimation error
has been obtained in the fully Bayesian context of minimum mean squared error. 
I have shown, in particular, that the Shannon equivocation, $h(X|Y)$, in the differential
sense cannot exceed $(1/2)\, \ln\,(2\pi e\,\MMSE)$, and hence the MI is bounded below by 
$h(X)-(1/2)\, \ln\, (2\pi e\,\MMSE)$. 

Both these results were generalized to the case of MIMO channels. However, the MMSE-based
lower bound on MI is not easily extendable to the discrete case. (I exclude here 
the trivial construct of associating with the PDF $P(x)$ of a continuous random parameter $X$ a
discrete PD involving probabilities $\{p_i\equiv P(x_i)\,\Delta x\}$ computed for finite bins, 
centered at regularly spaced points $x_i$ that are separated by an interval $\Delta x$ 
small compared to the scale over which $P(x)$ varies significantly.)

If additional input variables other than those of interest to the estimation problem
are present, in general they serve to compromise the fidelity with which the 
variables of interest may be estimated.
The impact of such nuisance variables on estimator performance was elucidated here with formulations based 
separately on MI, FI, and MMSE,
and a number of important inequalities were derived that provide valuable insight into
information and error-based metrics of performance.
The MMSE based description of the nuisance is particularly intriguing since 
it seems to predict a nearly counter-intuitive result that the presence of nuisance,
can never improve performance, even when it is
strongly coupled to the input and has vanishing variance, {\em i.e.,} independent
of its statistical correlations with the input.
This may be a peculiarity of how MMSE is defined, but surely deserves additional consideration.

\section*{Acknowledgments}
The author is pleased to acknowledge helpful contributions from S. Narravula.
Funding support from the Air Force Office of Scientific Research under grants
FA9550-08-1-0151 and FA9550-09-1-0495 is gratefully acknowledged.


\begin{thebibliography}{99}
\bibitem{Shannon48}
C.~Shannon, ``A mathematical theory of communication," Bell Syst. Tech. J., {\bf 27},
pp. 379-423 and 623-656 (1948).
\bibitem{CoverThomas91}
T.~Cover and J.~Thomas, {\em Elements of Information Theory}, Wiley (New York, 1991).
\bibitem{vanTrees68}
H.~Van Trees, {\em Detection, Estimation, and Modulation Theory}, Wiley (New York, 1968).
\bibitem{ClarkeBarron90}
B. Clarke and A. Barron, ``Information-theoretic asymptotics of Bayes methods," IEEE Trans. Inform. Th.,
{\bf 36}, pp. 453-471 (1990).
\bibitem{Rissanen96}
J. Rissanen, ``Fisher information and stochastic complexity," IEEE Trans. Inform. Th.,
{\bf 42}, pp. 40-47 (1996).
\bibitem{BrunelNadal98}
N. Brunel and J.-P. Nadal, ``Mutual information, Fisher information, and population coding,"
Neural Computation, {\bf 10}, pp. 1731-1757 (1998).
\bibitem{KangSompolinsky01}
K. Kang and H. Sompolinsky, ``Mutual information of population codes and distance measures in probability
space," Phys. Rev. Lett., {\bf 86}, pp. 4958-4961 (2001).
\bibitem{ChallisYarrowSeries08}
E. Challis, S. Yarrow, and P. Seri\`es, ``Fisher vs Shannon information in populations of neurons,"
preprint (2008). 
\bibitem{Duncan70}
T. Duncan, ``On the calculation of mutual information," SIAM J. Appl. Math. {\bf 19},
pp. 215-220 (1970).
\bibitem{GuoShamaiVerdu05} 
D. Guo, S. Shamai, and S. Verdu, ``Mutual information and minimum mean-square error in Gaussian channels,"
IEEE Trans. Inform. Th. {\bf 51}, pp. 1261-1282 (2005).
\bibitem{WolfZakai07}
E. Mayer-Wolf and M. Zakai, ``Some relations between mutual information and estmation error in Wiener space,"
Annals Appl. Prob. {\bf 17}, pp. 1102-1116 (2007).
\bibitem{footnote1} Whenever possible, a random variable is denoted by an upper-case roman letter and the values 
it may take by the corresponding lower-case letter. Also, for notational simplicity the same function 
label $P$ is used for the PDFs of different variables, even though the PDFs have, in general, 
different functional dependences on their arguments.
\bibitem{ChR51} D. Chapman and H. Robbins, ``Minimum variance estimation without regularity assumptions,"
Ann. Math. Stat. {\bf 22}, pp. 581-586 (1951).
\bibitem{Kay93} S. Kay, {\em Fundamentals of Statistical Signal Processing: Estimation Theory},
Prentice Hall (New Jersey,1993), Chapters 10 and 11.
\bibitem{Rao73}
C.~Rao, {\em Linear Statistical Inference and Its Applications}, Wiley (New York, 1973).
\bibitem{DMR94} A single-parameter, non-Bayesian version of this result was proved in 
A. D'Andrea, U. Mengali, and R. Reggiannini, ``The modified Cram\'er-Rao bound
and its application to synchronization problems," IEEE Trans. Commun., {\bf 42}, pp. 1391-1399 (1994).
\end{thebibliography}
\end{document}